\documentclass[10pt,preprint]{aastex}
\usepackage{fancyhdr,graphics}
\usepackage[colorlinks=true, urlcolor=blue]{hyperref}   
\usepackage{natbib}

\title{Measuring and Correcting Wind-Induced Pointing Errors of the
  Green Bank Telescope Using an Optical Quadrant Detector}

\author{Paul Ries\altaffilmark{1,2}, Todd R. Hunter\altaffilmark{1}, 
Kim T. Constantikes\altaffilmark{3}, Joseph J. Brandt\altaffilmark{4}, 
Frank D. Ghigo\altaffilmark{4}, Brian S. Mason\altaffilmark{1},
Richard M. Prestage\altaffilmark{4}, 
Jason Ray\altaffilmark{4}, Frederic R. Schwab\altaffilmark{1}}

\email{par9r@virginia.edu}
\altaffiltext{1}{NRAO, 520 Edgemont Rd, Charlottesville, VA, 22903} 
\altaffiltext{2}{University of Virginia, Astronomy Dept., P.O. Box 400325, 
Charlottesville, VA, 22904-4325} 
\altaffiltext{3}{Elbit Systems of America, Merrimack, NH, 03054}
\altaffiltext{4}{NRAO, Rt 28/92, Green Bank, WV, 24944} 

\begin{document}

\begin{abstract}
Wind-induced pointing errors are a serious concern for large-aperture high-frequency radio telescopes.  In this paper, we describe the implementation of an optical quadrant detector instrument that can detect and provide a correction signal for wind-induced pointing errors on the 100m diameter Green Bank Telescope (GBT).  The instrument was calibrated using a combination of astronomical measurements and metrology.  We find that the main wind-induced pointing errors on time scales of minutes are caused by the feedarm being blown along the direction of the wind vector. We also find that wind-induced structural excitation is virtually non-existent.  We have implemented offline software to apply pointing corrections to the data from imaging instruments such as the MUSTANG 3.3~mm bolometer array, which can recover $\sim 70$\% of sensitivity lost due to wind-induced pointing errors. We have also performed preliminary tests that show great promise for correcting these pointing errors in real-time using the telescope's subreflector servo system in combination with the quadrant detector signal.
\end{abstract}


\section{Introduction}

\begin{figure}[H]
\centering
\resizebox{75mm}{!}{\plotone{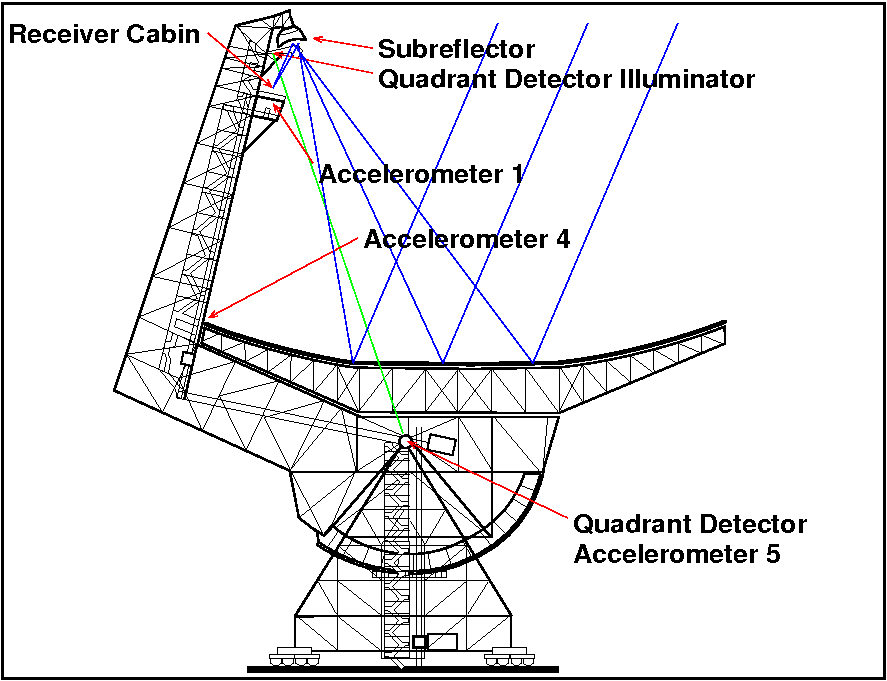}}
\caption{Layout of the GBT shown in profile observing at an elevation
of $66\arcdeg$.  Blue lines show the approximate optical path of the
telescope for high-frequency instruments. The green line shows the
optical path between the illuminator and the quadrant detector.}
\label{fig:gbtlayout}
\end{figure}

Located at an elevation of 822m in Green Bank, West Virginia, the
Green Bank Telescope (GBT) \citep{prestage2009} is a 100m fully steerable
single-dish radio telescope.  The optical design is a dual offset
Gregorian on an alt-az mount providing an unblocked aperture of 100m
diameter.  The primary paraboloid is formed by a segmented active
surface of 2004 panels \citep{Lacasse1998} which enables observations
from frequencies of 300 MHz to 115 GHz.  The high-frequency receivers
($\nu > 1.15$~GHz) illuminate the concave ellipsoidal subreflector
from a receiver cabin at the Gregorian focus \citep{gbt2004}.
Figure~\ref{fig:gbtlayout} shows a profile diagram of the telescope.
At high frequencies (above 50~GHz), the diffraction beam size of the
GBT ($\approx \frac{740''}{\nu (\mathrm{GHz})}$) is less than $15''$, so
observations become particularly sensitive to uncorrected pointing and
tracking errors.  Indeed, the two primary telescope-related
performance factors that limit high-frequency observing are pointing
accuracy and surface accuracy.

In order to improve the GBT pointing and surface performance, the
Precision Telescope Control System (PTCS) project was initiated in
early 2003.  It is well-known that changes in the thermal environment
of large radio telescopes have a considerable effect on the pointing
\citep{bayley1994}.  Therefore, the first major PTCS deliverable was
the implementation of a structural temperature network and
inclinometer system to allow dynamic corrections to augment the static
pointing model based on real-time measurements
\citep{constantikes2004,ptcs2004}.  Following the recent upgrade of
the telescope azimuth track \citep{symmes2008}, this system has
produced an all-sky blind pointing performance in low winds ($<3$~m~s$^{-1}$)
of $5''$ rms, or about half the diffraction beamsize at 90~GHz
\citep{constantikes2007,prestage2009}.  The second major PTCS effort
was the application of the technique of out-of-focus (OOF) holography
\citep{nikolic2007a} which succeeded in measuring the large-scale
gravitational deformation of the surface that remains after
application of the finite-element model. This technique yielded a set
of elevation-dependent Zernike coefficients (through fifth order)
which are now automatically applied and produce a nearly constant
telescope gain vs.\ elevation during nighttime observations
\citep{nikolic2007b}.  In 2008, the OOF measurement and active surface
correction sequence was automated to allow fine tuning of the surface
collimation with minimal overhead, thereby expanding the possibility
of high-frequency daytime observations during which the telescope
thermal environment can change quickly \citep{hunter2009}.  Finally,
in 2009 the small-scale rms surface error of the GBT was reduced
significantly (from 390 to 240 microns rms) by refining the 2,209
actuator home positions during a campaign of traditional radio
holographic measurements using a 12~GHz satellite beacon
\citep{holo2010}. This improvement in surface accuracy resulted in a
240\% increase in the 90~GHz efficiency (Hunter et al., in
preparation).

Following all of the recent surface improvements, the focus of the
PTCS project activity has shifted back to pointing performance.  The
efforts described in this paper center on measuring and reducing the
dynamic pointing error.  On the GBT, it is motion of the vertical
feedarm that is the largest source of dynamic pointing errors, which
we define as those that can occur over the course of observing a
single source, i.e. on timescales of seconds to minutes
\citep{smith2000}.  Although the GBT feedarm design provides an
unblocked aperture, it is less stiff than a traditional on-axis
quadrupod design. Motion of the feedarm tip results in a relative
movement of the subreflector with respect to the primary mirror which
appears (to first order) as a pointing error in the focal plane.
These deflections can become significant when the feedarm undergoes
structural resonance or when the feedarm is buffeted by winds.  The
wind problem is especially acute for high-frequency observing.  High
frequency performance depends on good atmospheric transparency and
good pointing.  At wind speeds above $\sim$4~m~s$^{-1}$, the
feedarm deflections become the dominant constraint on the
high-frequency performance of the telescope.  Reducing the
wind-induced pointing errors could nearly double the amount of
available high-frequency observing time because periods of good
transparency can sometimes occur accompanied by significant winds,
particularly in winter.

\section{Instrument}

\subsection{Description} 

The approach we have taken to investigate dynamic pointing errors on the GBT is to add a simple feedarm metrology system to the telescope. Although the GBT pointing performance was originally designed to utilize corrections from a complex laser metrology network \citep[see e.g.][]{parker99}, development work on this network was postponed in late 2003 in order to concentrate resources on improving the thermal pointing model and pursuing the more conventional techniques of surface holography.  However, by this time, one component of the optical metrology scheme had already reached an advanced stage. This device consists of an optical illuminator aimed at a position sensitive detector (PSD).  For historical reasons, we traditionally refer to the aggregate system as a ``quadrant detector'' (QD) as it was originally conceived by John M. Payne to operate in this manner in the early 1990's \citep{Hall1998}.  QDs (and more generally PSDs) have had many useful applications in astronomy, such as for optical image stabilization \citep{stilburn1997,stilburn1992} and X-ray astronomy \citep{culhane1991}.  In our case, the purpose of the QD is to provide two-dimensional positioning information of the feedarm relative to the telescope elevation axle in approximately the same plane as the focal plane of the telescope.  Thus, if properly calibrated, its datastream can be used to detect and correct for wind- or resonance-induced pointing errors from the feedarm incurred during astronomical observations.

\begin{figure}[H]
\centering
\plottwo{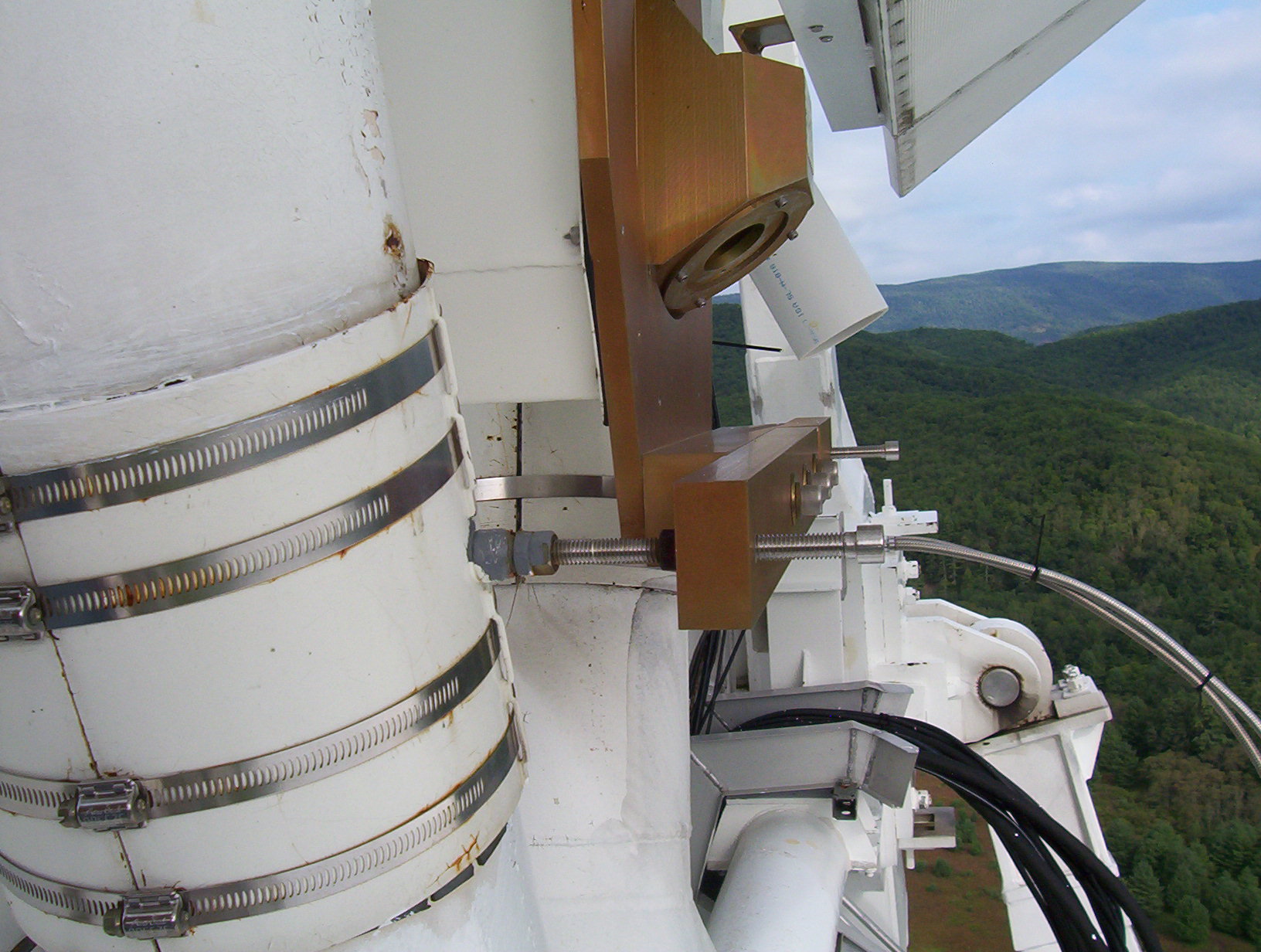}{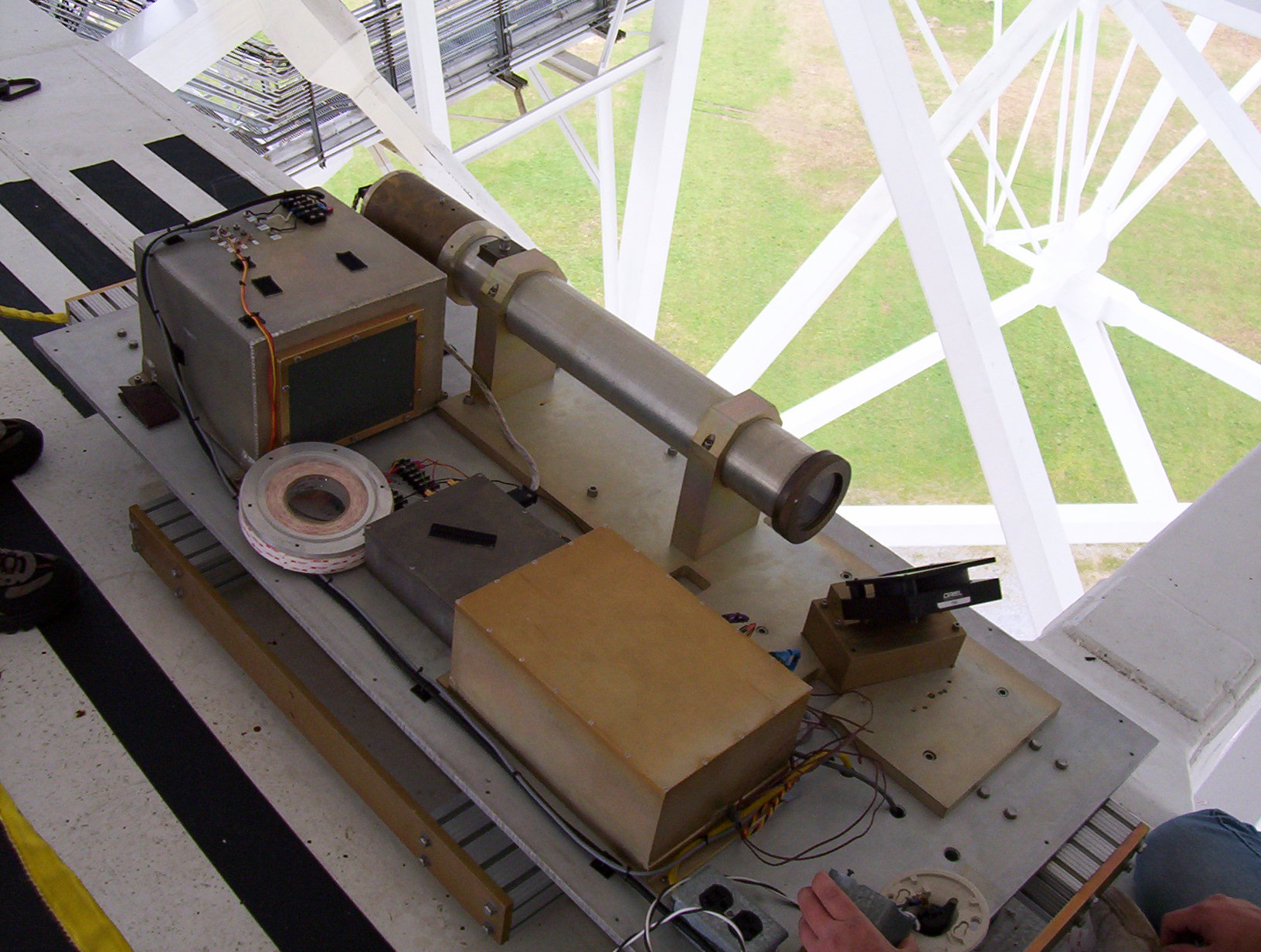}
\caption{Left panel) Photo of the QD illuminator assembly (gold
anodized housing) located just below the subreflector; right panel) 
Photo of the QD telescope tube, mirror, and electronics with the
weather cover removed.  It is located above the elevation 
axle and about 15m under the primary dish.}
\label{fig:photos}
\end{figure}

After testing several generations of prototypes, the current version of the QD was deployed in late 2004. Figure~\ref{fig:gbtlayout} shows the location of the illuminator and detector with respect to the general telescope layout.  The illuminator assembly is located near the tip of the vertical feedarm, slightly below the subreflector (Figure~\ref{fig:photos}).  The illuminator source is a Luxeon III Star 1400~mA green LED. A 150~mm focal length lens produces a slightly divergent beam in order to maintain illumination of the detector at all possible telescope orientations.  A 12-inch hole in one panel of the dish allows light from the illuminator to reach the detector assembly underneath. The hole lies about 10m to the right of the symmetry axis of the telescope, which allows the detector assembly to be placed on an accessible platform near the right elevation bearing. Inside the weather-tight detector box is a flat turning mirror followed by a telescope with an 800~mm focal length lens that focuses the beam onto a 4~mm square duolateral PSD (Figure~\ref{fig:photos}).  The total path length of the beam is $\approx 87$~m.  The PSD produces four voltages, each of which is amplified, filtered by an active filter centered at the modulation frequency, and then detected by an RMS-to-DC converter.  The converter output voltages are digitized at $\approx 8.8$~Hz and timestamps are attached by a  microcontroller.  The digitized signal is then transmitted via an RS-232 serial link to a concentrator, which then sends the signal over ethernet  back to the control room. 

\subsection{Data products} 

The centroid of the illuminator spot on the surface of the detector is given by the unitless ratio of the differences between the two pairs of PSD output signals to their sum (i.e. $\mathrm{QD}_X=\frac{V_1-V_2}{V_1+V_2}$, $\mathrm{QD}_Y=\frac{V_3-V_4}{V_3+V_4}$).  The PSD is large enough that the
illuminator spot is fully captured for all elevations above $15\arcdeg$, but the spot begins to walk off the PSD below that point. This limitation is acceptable because high-frequency observations are not generally attempted at such low elevations due to atmospheric opacity.  The GBT monitor and control (M\&C) system collects and stores the QD data into binary FITS tables \citep{cotton1995} along with several calibrated data products derived from these voltages (described later).  A plug-in tool is available in ASTRID, the GBT user interface software \citep{oneil2006}, to graphically display the QD-inferred pointing error in real-time. The running rms of the data is also displayed in text format on the GBT status window which is available to GBT observers and operators in ASTRID.  The illuminator and detector assemblies each contain a number of sensors that measure
the temperature of the internal air, the electronics, and the optics, which are also timestamped and stored in a separate binary FITS file.

\subsection{Calibration}
\subsubsection{Overview}

The first step in the calibration process was to remove the signature of gravitational deformation of the feedarm from the QD signal, and we did so by fitting the raw QD signal at many elevations with a model (described in the next section).  The next step was to calibrate from raw QD coordinates to a pointing error.  We assumed that the physics were linear and described it in the form of a matrix equation:

\begin{equation}
\underbrace{
    \left| \begin{array}{cc}
      D_{11} & D_{12}  \\
      D_{21} & D_{22}  
    \end{array} \right|}_D
\underbrace{
    \left| \begin{array}{cc}
      QD_{X,cor} & \cdots \\
      QD_{Y,cor} & \cdots 
    \end{array} \right|}_{p_{corr}}
=
\underbrace{
    \left| \begin{array}{cc}
       \delta \phi& \cdots \\
       \delta \theta & \cdots 
    \end{array} \right|}_{err}
\end{equation}
where $\delta \phi$ is the offset in elevation and $\delta \theta$ is the offset in cross-elevation and the QD variables are the QD X and Y  after the removal of the signal of gravitational deformation.  Mathematically, the 2x2 matrix represents a combination of rotations and scalings required to convert the raw QD signal with the gravitational deflection removed to pointing errors in elevation and cross-elevation.  However, the elevation and cross-elevation calibrations in this formulation are independent of each other.  Therefore, we calibrated them using different methods: elevation using a conversion between the surveyed deflection of the feedarm with elevation, and the QD measured motion of the feedarm with respect to elevation and cross-elevation using a direct conversion between astronomical pointing measurements and QD measured pointing errors.  The final result of our calibration is given below in units of arcsec:

\begin{equation}
\left| \begin{array}{ccc}
      D_{11} & D_{12} & \\
      D_{21} & D_{22} & 
    \end{array} \right|
= 
\left| \begin{array}{ccc}
      -550.9 & -156.4 & \\
      +570.0 & -468 & 
    \end{array} \right|
\end{equation}

\subsubsection{Removal of Gravitational Deflection}

\begin{figure}[H]
\centering 
\resizebox{75mm}{!}{\plotone{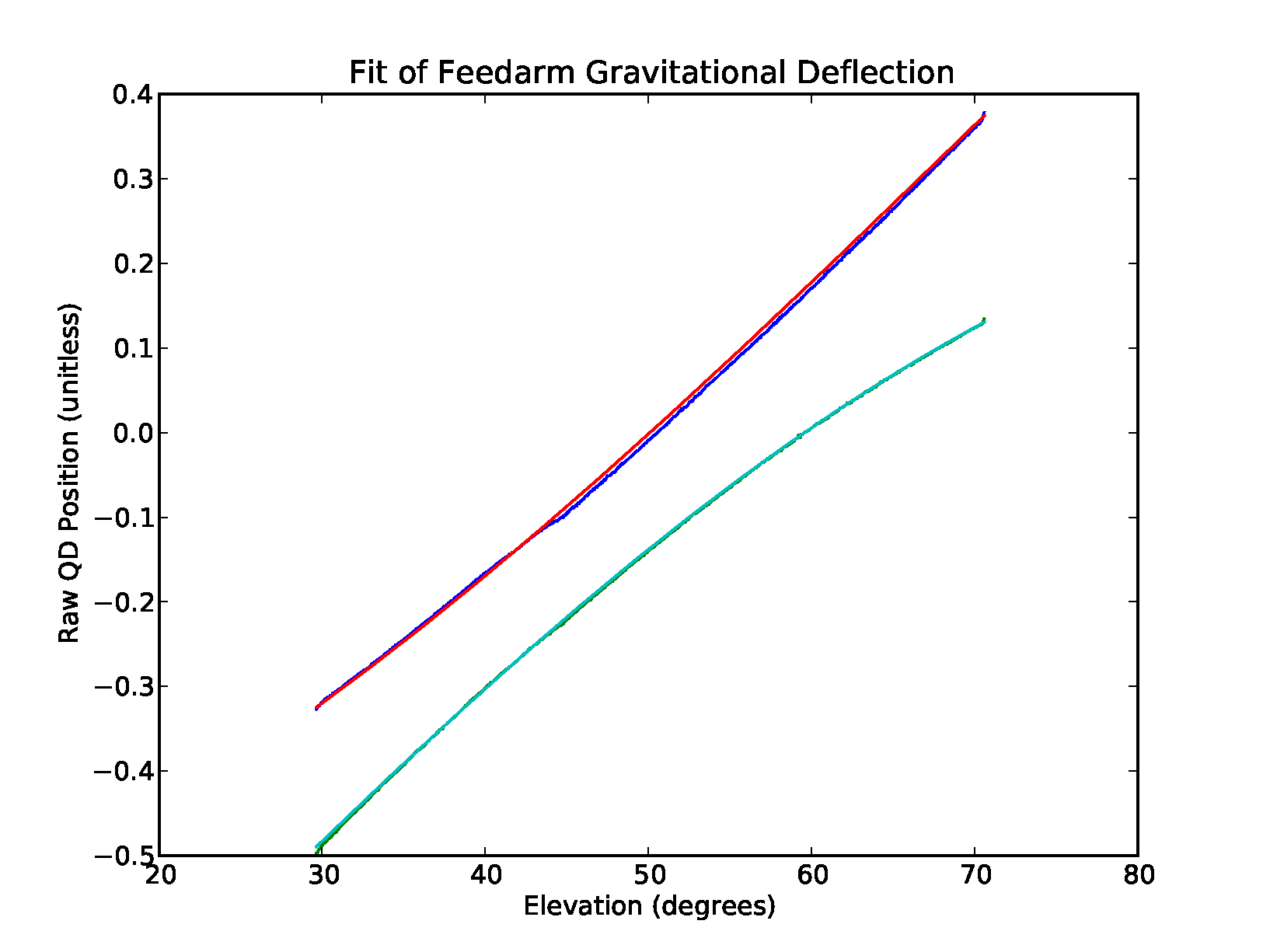}}
\caption{Signal of the feedarm deformation (blue and green) due to the
change in the gravity vector at different elevations, along with the model
fit (red and cyan).  This deformation is already accounted for in the
telescope pointing model, so it must be removed from the QD signal in
order to isolate the residual dynamic pointing errors that it also
measures.}
\label{fig:gravdef}
\end{figure}

The existing static pointing model for the telescope already accounts for the gravitational deflection of the feedarm, so the deflection must be removed from the QD data in order to isolate the residual dynamic pointing errors.  This problem was fairly trivial to resolve.  We simply tipped the telescope from approximately 70 to 30$\arcdeg$ elevation and measured the resulting spot position on the QD (referred to as QD$_{X}$ and QD$_{Y}$, which can range from a unitless -1 to +1).  We assumed that the deflection was linear in trigonometric functions of elevation ($\phi$), i.e. of the form:

\begin{equation}
 QD_{\mathrm{gravdef}} = A\cos(\phi)+B\sin(\phi)+C
\end{equation}

and then simply performed a linear least squares fit for the coefficients based on the thousands of data points acquired during a single tip of the telescope.  The gravitational deformation is then removed using the equation:

\begin{equation}
\left| QD_{\mathrm{corr}}\right| = QD_{\mathrm{obs}} - (A\cos(\phi)+B\sin(\phi)+C)
\end{equation}

for each axis of the telescope, with different coefficients for each axis.  Figure \ref{fig:gravdef} shows the results from the tip and the least squares fit.  What remains after this subtraction is the relative position of the spot on the QD with respect to the standard ``no error'' position.  

\subsubsection{Elevation Calibration: Tipping Survey}

\begin{figure}[H]
 \centering
\resizebox{75mm}{!}{\plotone{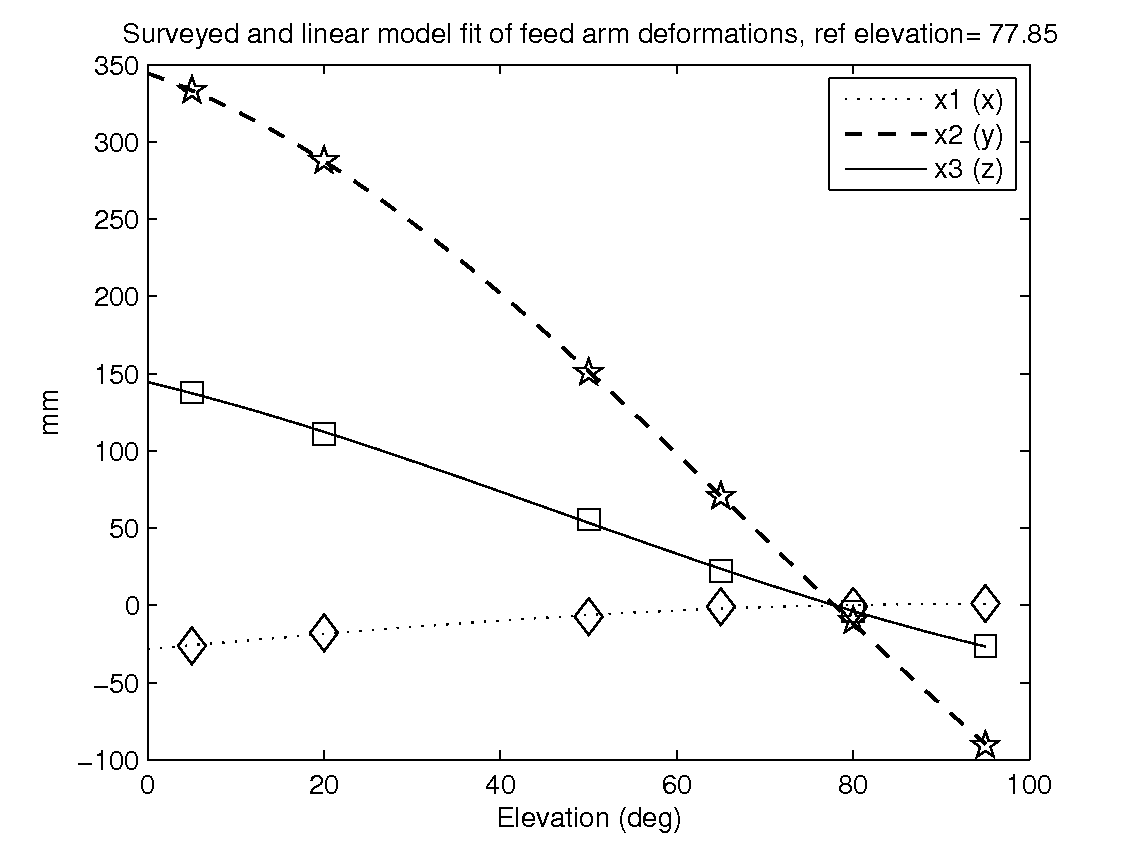}}
\caption{The relative deflection of the GBT feedarm vs.\ elevation as
  measured by traditional surveying techniques.  The points are the
  measurements and the lines are the best fits.}
\label{fig:kimcalib}
\end{figure}

As has been stated previously, the feedarm deflects due to the changing gravity vector.  In order to calibrate the QD {\it in situ}, we consulted archival data from an elevation tipping survey of the GBT performed on September 7, 2000. This survey used traditional techniques employing a Leica total station instrument to determine the change in position of the feedarm vs.\ elevation with respect to a fixed point.  Measurements were recorded at 6 different elevations from $5\arcdeg$ to $95\arcdeg$ and resulting deflections were calculated. A fit to the measured deflection of the feedarm location of the illuminator with respect to the detector can be seen in Figure~\ref{fig:kimcalib}.  This fit gave us a conversion between raw QD signal and the deflection in millimeters of the feedarm.  We then calculated the conversion from millimeters of motion of the feedarm to resulting pointing error on the sky simply based on the geometry of the telescope optics.  Combining these two steps gave us the conversion between raw QD signal and pointing error.

Unfortunately, this method has a major drawback: the displacement of the feedarm due to an elevation change is primarily in the elevation direction.  Since the cross-elevation direction is, by definition, perpendicular to the gravity vector at all elevations, the feedarm does not change in position substantially in cross-elevation as the telescope moves in elevation.  As a result, the tipping survey leads to a poor fit in cross-elevation since there is almost no change in cross-elevation to fit.

\subsubsection{Cross-Elevation Calibration: Astronomical Observations}

An alternate method of calibrating the QD {\it in situ} is to compare it
directly to the result of ``half-power tracks'', which measure the
dynamic pointing of the telescope.  In a half-power track, one
observes a bright point source at the half-power point of the telescope
beam, offset in either the elevation or cross-elevation directions.
The slope of the receiver's beamshape (which is effectively Gaussian
due to the typical 14~dB edge tapered illumination pattern of the feedhorn)
at the half-power point gives the derivative of the antenna
temperature (i.e. received power) with respect to the change in
pointing offset.  This slope can then be used to convert the changes
in antenna temperature during a track at the half-power point of the
beam into pointing error in arcseconds vs.\ time.  By using the slope at
the half-power point as the conversion factor, we get a pointing error accurate to within 2\%
over a range of $\pm$ $10''$ from the half-power point at Ku-band
(14~GHz) where the beamsize is $53''$.  Atmospheric variations can
also cause fluctuations in antenna temperature, however we can remove
most of these fluctuations by differencing the source signal with the
signal from the second receiver beam which is offset by $330''$ in
cross-elevation and observing blank sky.  As a result, we were able to
accurately measure true pointing errors of up to $10''$ during the
half-power tracks.  To extract larger pointing errors, a more rigorous
analysis should use the inverse of the error function
\citep{smith2000}; however, our simpler approach was sufficient for our needs.

\begin{figure}[H]
\centering
\plottwo{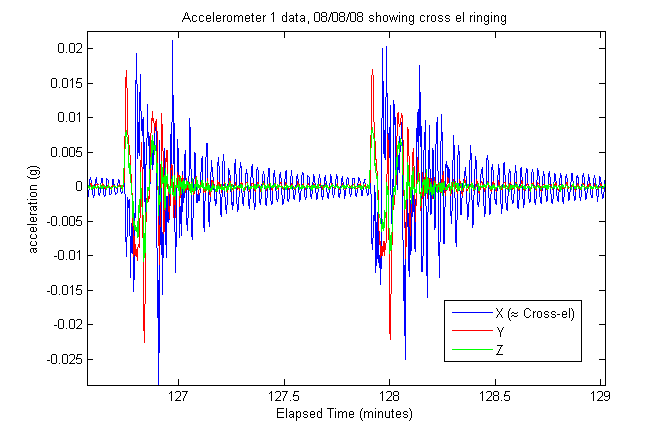}{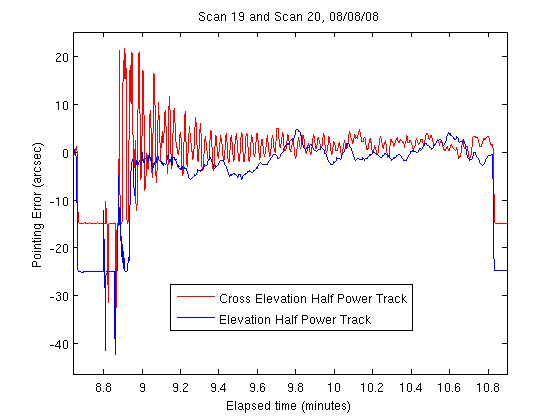}
\caption{Data from both the three-axis accelerometer (left) in units of g-force
  and from half-power tracks (right) demonstrating that feedarm oscillation
  occurs almost entirely in the cross-elevation direction.}
\label{fig:elvscel}
\end{figure}

By analyzing the results of dozens of half-power tracks under different observing conditions, we gathered some general conclusions. Under benign (low-wind, long sidereal tracks on a source or its half-power point) conditions, feedarm motion is not the dominant source of pointing error.  We found two conditions under which feedarm motion can dominate: when winds are $>\sim$4-5 m~s$^{-1}$ or when structural oscillations are induced in the feedarm by the telescope slews.  Because these oscillations decay quickly ($\approx 60$~s in duration), we modified the half-power track technique to include a periodic series of ``kicks'' using a custom trajectory.  Kicks consist simply of moving the telescope off-source by a substantial fraction of a degree, then back to the half-power point as rapidly as possible. By spacing the kicks at one minute intervals, the feedarm can be kept nearly constantly oscillating  during a half power track.  From the result of these tests, it is clear that the feedarm oscillates primarily in the cross elevation direction with very little effect seen in the elevation direction, such as can be seen in Figure~\ref{fig:elvscel}.  

A calibration was obtained by doing a direct least squares fit of the QD data to the pointing on the sky during these half-power tracks with kicks.  Since the excitation of the feedarm resonance is almost entirely in cross elevation, this method yielded a good and consistent fit in cross-elevation, but poor and inconsistent fits in elevation.  We therefore took the cross-elevation calibration from this method and combined it with the earlier elevation calibration from the tipping calibration in order to obtain our final overall calibration in both axes.

\subsection{Median subtraction}

Over the course of our experiments, we have discovered that, even after calibration, some residual errors exist in the QD signal due to drift in the zero point of the QD signal.  The most frequent cause of QD drift is thermal deformation in the QD assembly. A clear diurnal effect is frequently seen in the signal, with the steepest slope of order 4-10 arcsec per hour during the heating and cooling periods.  The signal is generally stable after midnight.  We solve this problem by taking a median of the QD data, either before or during the period of interest and subtract the median from the data. Following this correction, the QD data appear to be stable on at least 1 hour timescales, especially at night.  

We also occasionally get some hysteresis events in the QD which change the zero point by up to 150 arcseconds and are likely caused by oilcanning in the mount for the detector assembly.  However, these events are rare (one every few days) so median subtraction is adequate for removing them from the data.  Median subtraction also eliminates any problems of temporal variation in the total refraction along the optical path from the illuminator to the detector.  Median subtraction cannot, however, solve the problem of a loss of QD signal due to a fog, rain, or snow event interrupting the beam.  Therefore we simply do not use the QD when the weather is interfering the beam.

\section{Engineering Applcations}

\begin{figure}[H]
\centering
\plottwo{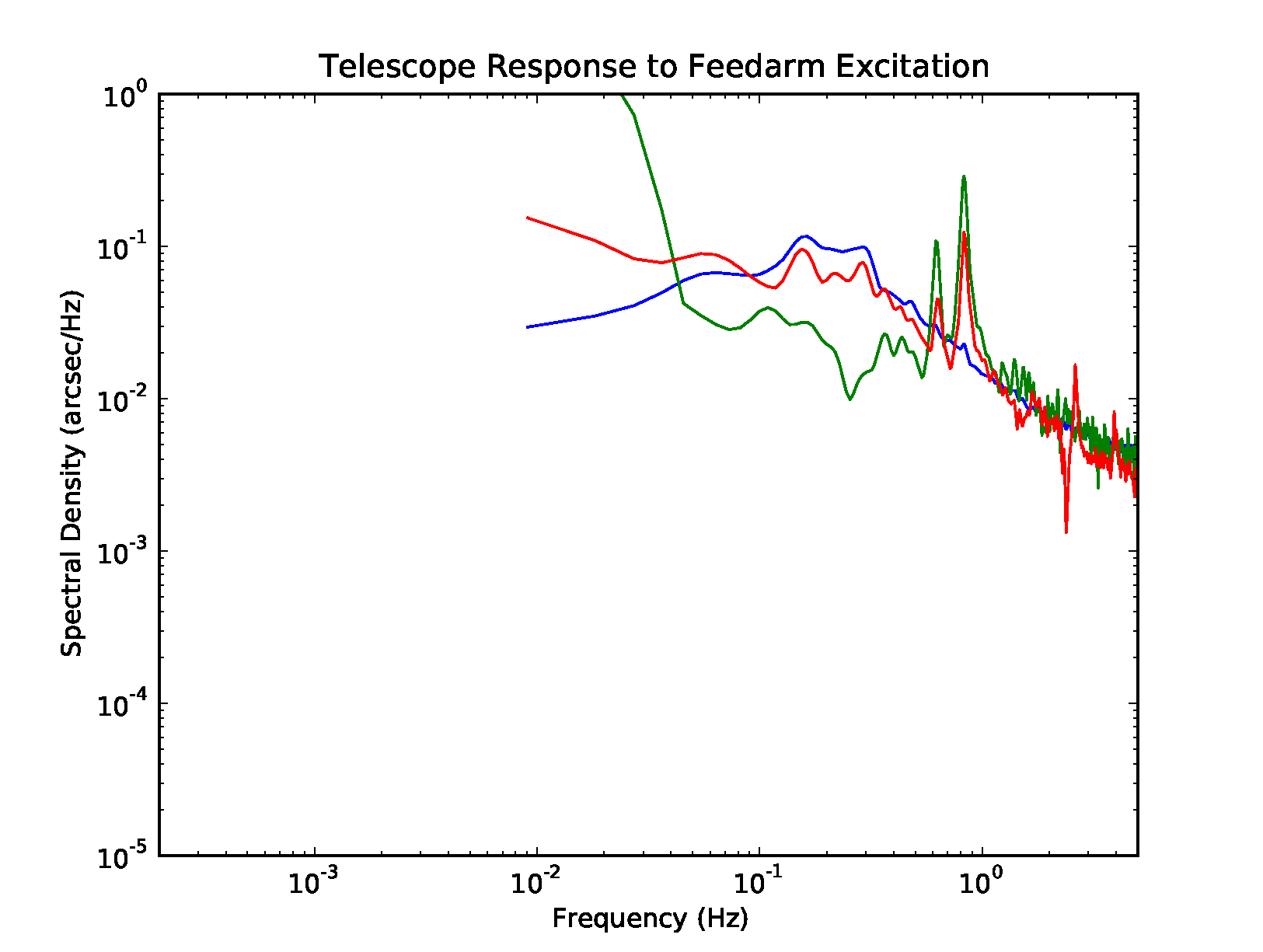}{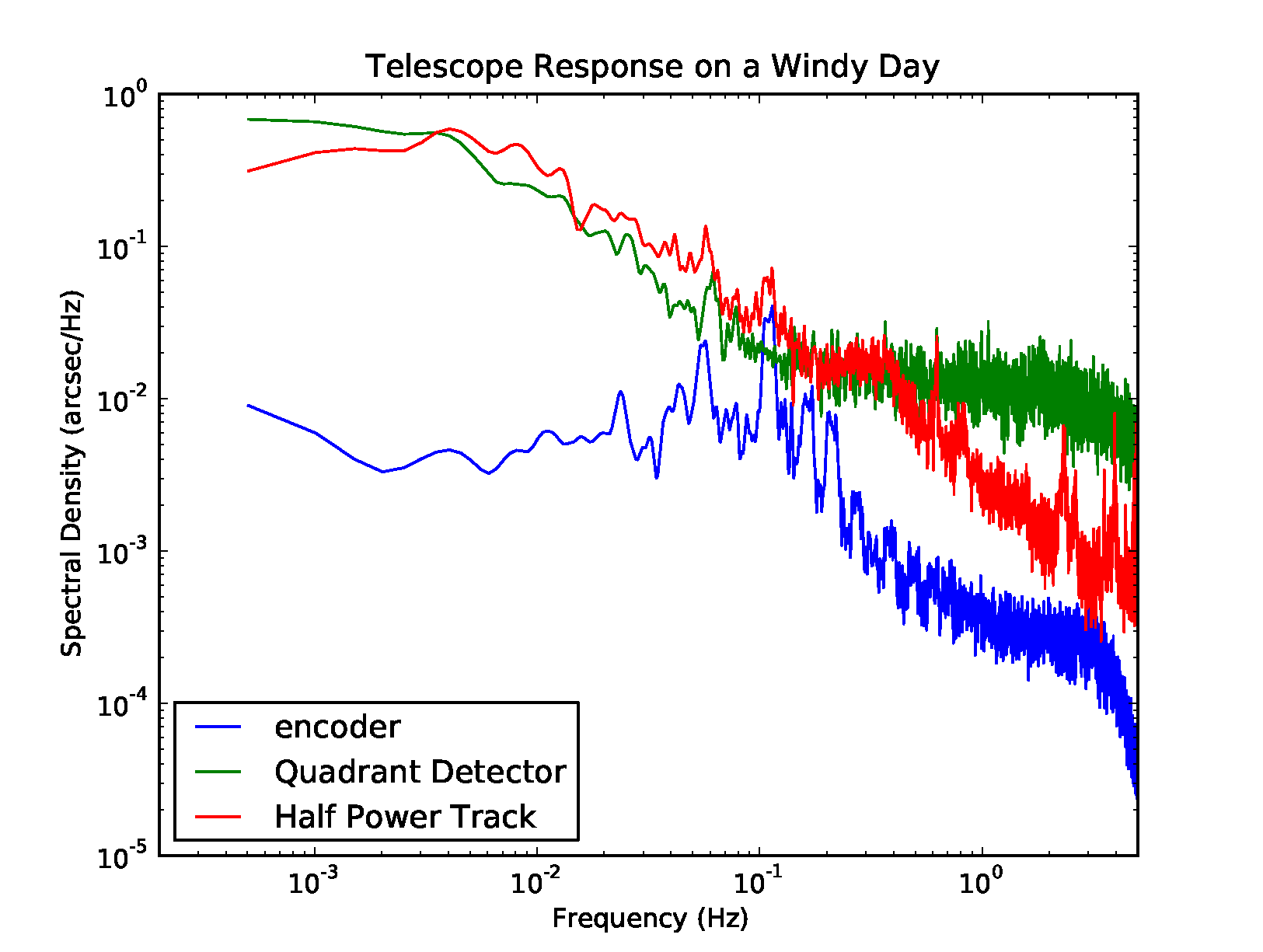}
\resizebox{75mm}{!}{\plotone{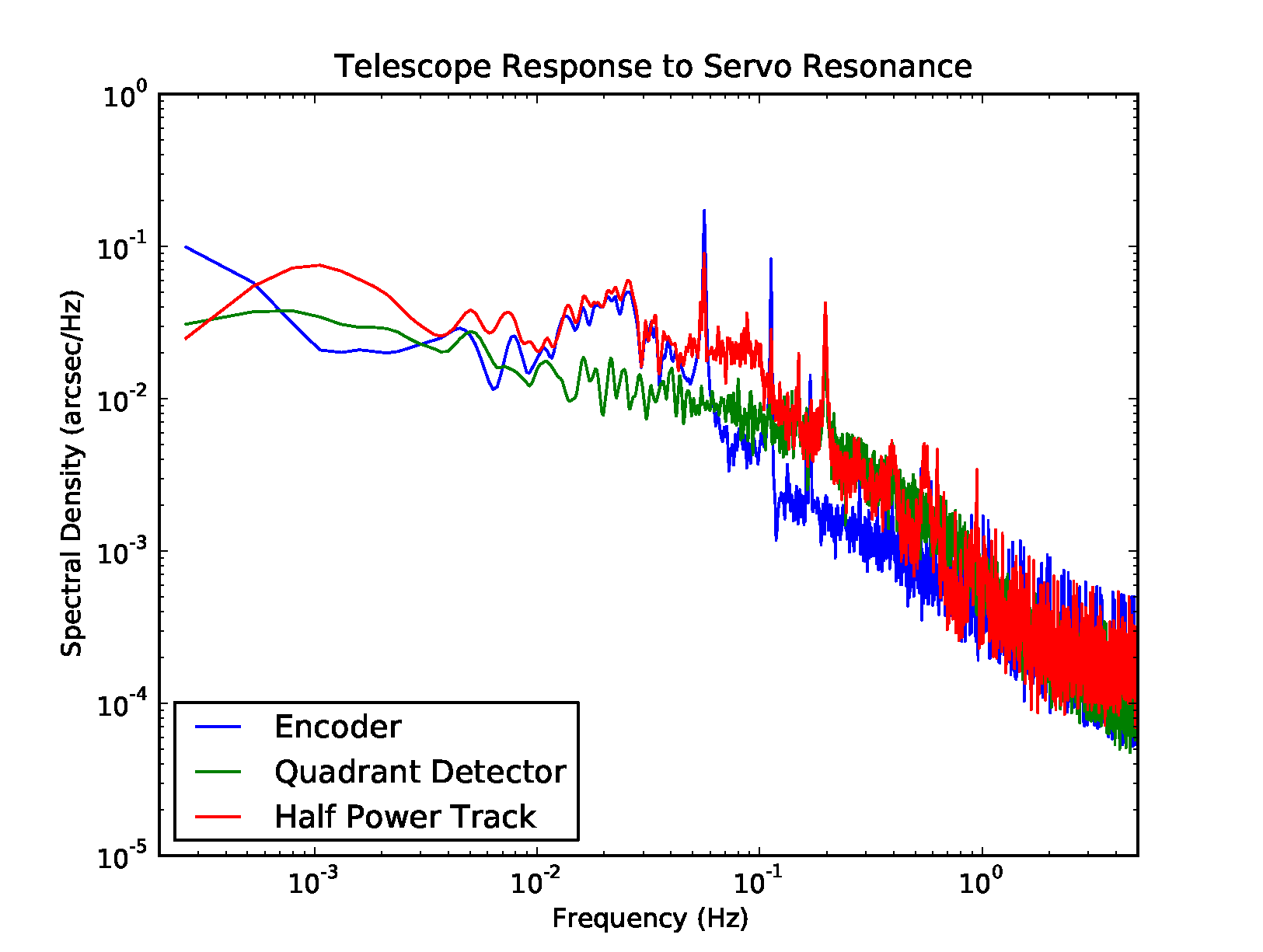}}
\caption{Frequency-space diagrams showing the magnitude of the
observed pointing errors (inferred from the receiver half-power track
data, the QD, and the encoder) vs.\ frequency under various different
conditions: upper left) feedarm resonance; upper right) windy
conditions; bottom) servo resonance.}
\label{fig:fftrifecta}
\end{figure}

\begin{figure}[H]
\centering
\plottwo{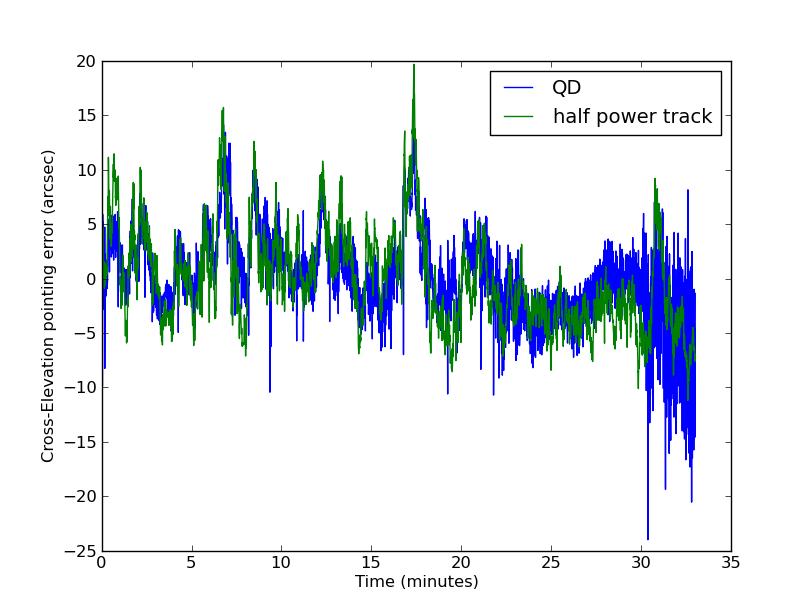}{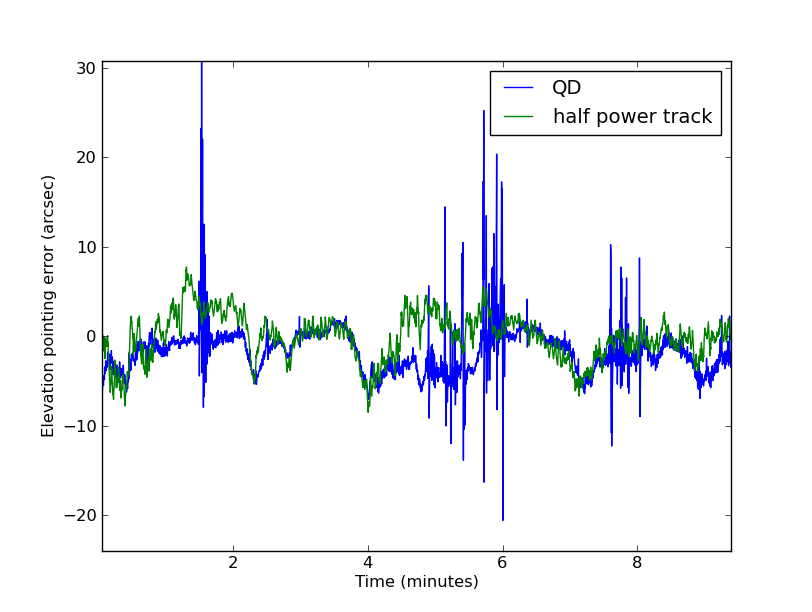}
\caption{Plot comparing the QD-inferred pointing error and half-power
track pointing errors in the time domain.  The average wind speed
during these observations was 3 m~s$^{-1}$.  Note that the cross-elevation
errors are larger and also yield a much better fit to the QD data.}
\label{fig:qdwindhp}
\end{figure}

We can use the calibrated QD to better study the nature of the dynamic pointing errors on the GBT and how they affect science observations.  There are three conditions on the telescope under which dynamic pointing errors are the dominant limit on sensitivity.  The first case is when a resonance in the servo control system is active, causing an oscillation in the tracking error, the second case is when the feedarm's structural resonance is excited, and the third case is when the feedarm is buffeted by winds.  The QD effectively detects most of the pointing errors in the latter two cases. Figure \ref{fig:fftrifecta} shows the Fourier transforms of pointing data from half-power tracks, the calibrated QD data, and the servo following error. Of these errors, the ones due to wind are the most interesting, as they are the largest and constrain high-frequency observing on the GBT. Prior to the use of the QD, it was often assumed that wind-induced pointing errors were due to excitation of the feedarm structural resonance \citep[see e.g.][]{wells2000} and would be correlated with wind speed.  Both of these assumptions, however, are challenged by the actual QD data.  One can see in Figure \ref{fig:fftrifecta} that the primary structural resonances of the feedarm at 0.6 and 0.8~Hz are barely visible above the continuum on a windy day.  Instead, wind-induced pointing errors take place on a timescale of minutes.  Figure \ref{fig:qdwindhp} shows the comparison of the QD data and two windy half-power tracks in the time domain, demonstrating that our calibration is reasonably accurate.

\begin{figure}[H]
\centering
\plottwo{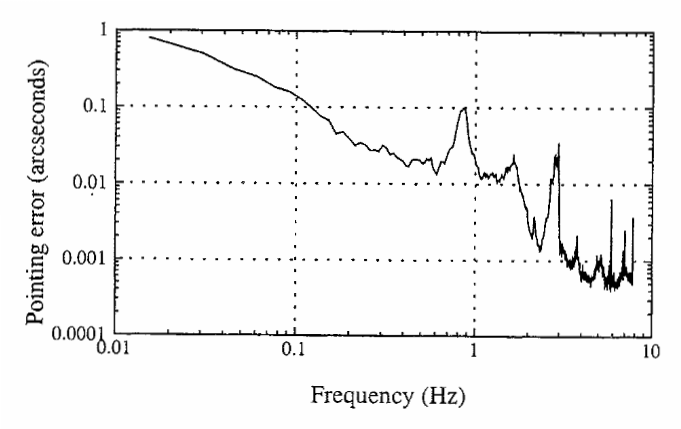}{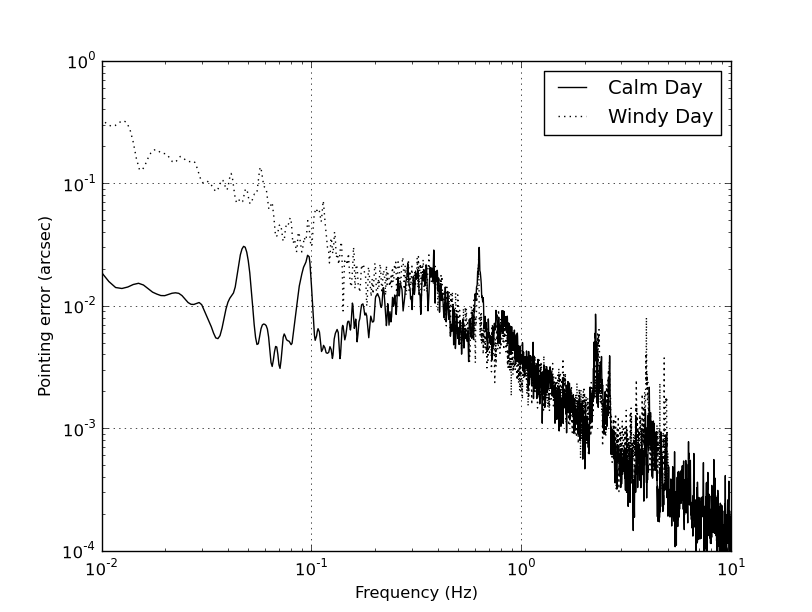}
\caption{Comparison of the Fourier transform of pointing errors
  between the Nobeyama 45~m telescope from \citet{smith2000} (left)
  and the GBT at Ka band (right).  The Nobeyama data and our windy
  data were both taken at wind speeds of $\approx$ 4~m~s$^{-1}$.  Our
  calm data comes from a day when wind speeds were $<$ 1~m~s$^{-1}$.
  There is no difference between the GBT structural resonances (0.6,
  0.8, and 2.1~Hz) on the calm day (February 28, 2008) and the windy
  day (January 24, 2008).  In fact, the only difference is the
  presence of more ``continuum'' power at frequencies under 0.3 Hz.
  The features at 0.046 and 0.093~Hz seen in the calm day's data are
  due to the known servo resonance and correspond to the azimuth motor
  axle rotation rate and its harmonic.  The exact frequency of these
  features depends linearly on the tracking velocity.  On the calm
  day, the scan length was 20 minutes with an average tracking rate of
  $3.69\arcsec$ sec$^{-1}$ while on the windy day the scan length was
  33 minutes with an average rate of $4.34 \arcsec$ sec$^{-1}$.  Thus,
  the spectral features appear at slightly higher frequencies in the
  windy day's data, but their relative contribution to pointing error 
  compared to the continuum is less.}
\label{fig:loglogshowdown}
\end{figure}

\citet{smith2000} performed similar measurements under windy conditions on the Nobeyama 45~m telescope.  Rather than half-power tracks, they used lunar limb scans to measure astronomical pointing, which serve similar purposes, but eliminate any directional ambiguity.  They found a similar result in that their largest pointing errors were from ``quasi-static'' motions (i.e. those with a frequency lower than 0.1 Hz).  However, they also found several structural resonances in their pointing data that were still significant and claimed that these resonances were excited by the wind.  Figure~\ref{fig:loglogshowdown} shows one of their lunar limb scans and two of our azimuth half-power tracks, including one under comparable conditions ($\approx 4$~m~s$^{-1}$~winds) and one under calm conditions with similar smoothing applied to the data. It is readily apparent in our data that the structural resonances are not excited at all by the wind.  In frequency space, the pointing behavior of the GBT above 0.3 Hz seems to be the same regardless of the wind. The presence of wind simply raises magnitude of the the low frequency ($<$0.3 Hz) pointing errors.  With the addition of QD data, we can say for certain that the low frequency errors we observe are due to the motion of the feedarm.  Additionally, our data extends over 20 and 33 minutes, rather than a few minutes for the Nobeyama data.  We continue to detect wind-induced low frequency pointing errors at even lower frequencies than were measured in the previous work, although the spectrum flattens below 0.01 Hz. This flattening is consistent with the typical spectral distribution of wind disturbances at 100m above the ground \citep[p. 61]{simiu1996}, but different from our 25m elevation observed wind spectrum, which follows a $\nu^{-1}$ power law and turns over at a frequency of about 0.001Hz.

\begin{figure}[H]
\centering
\plottwo{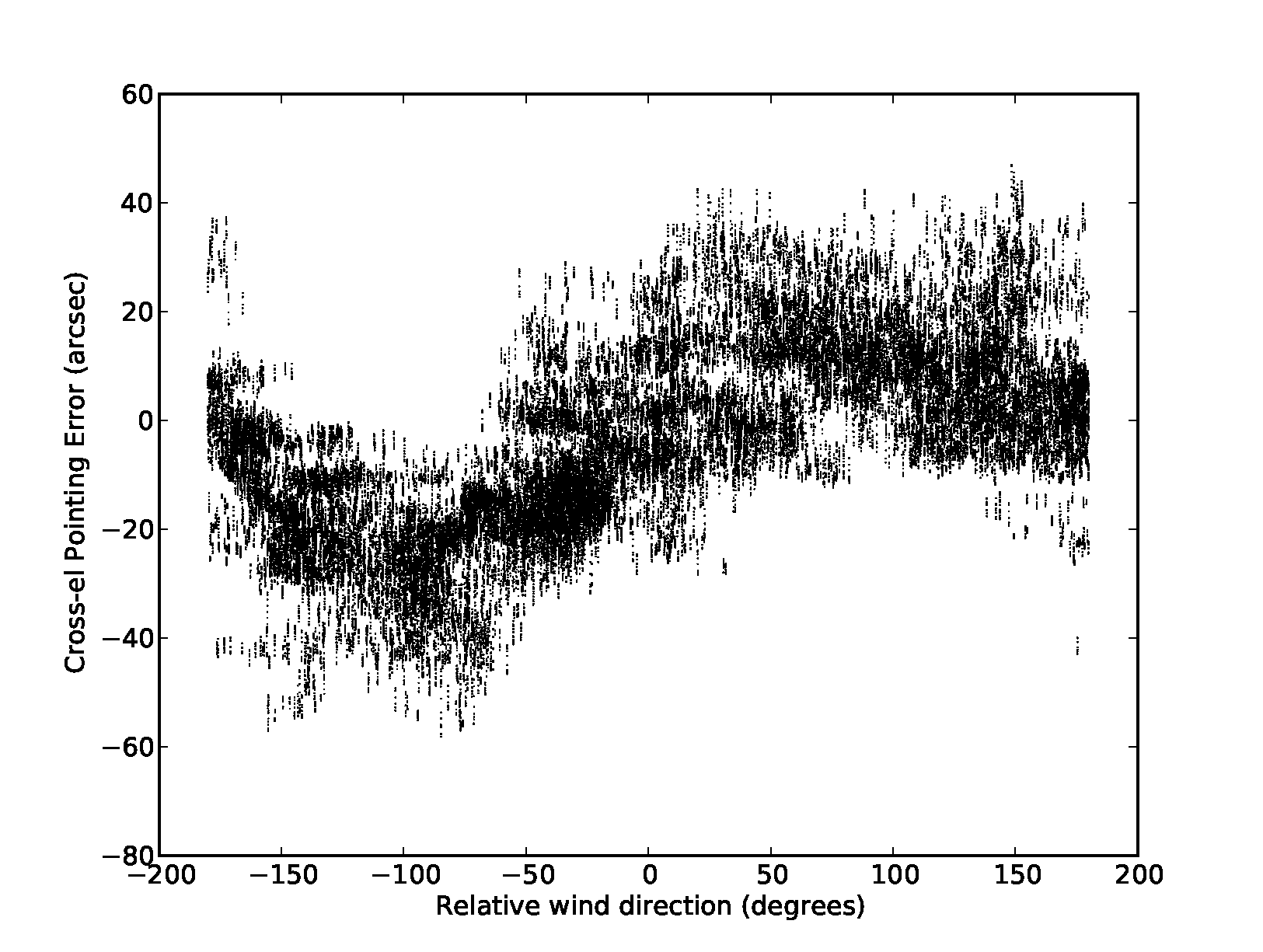}{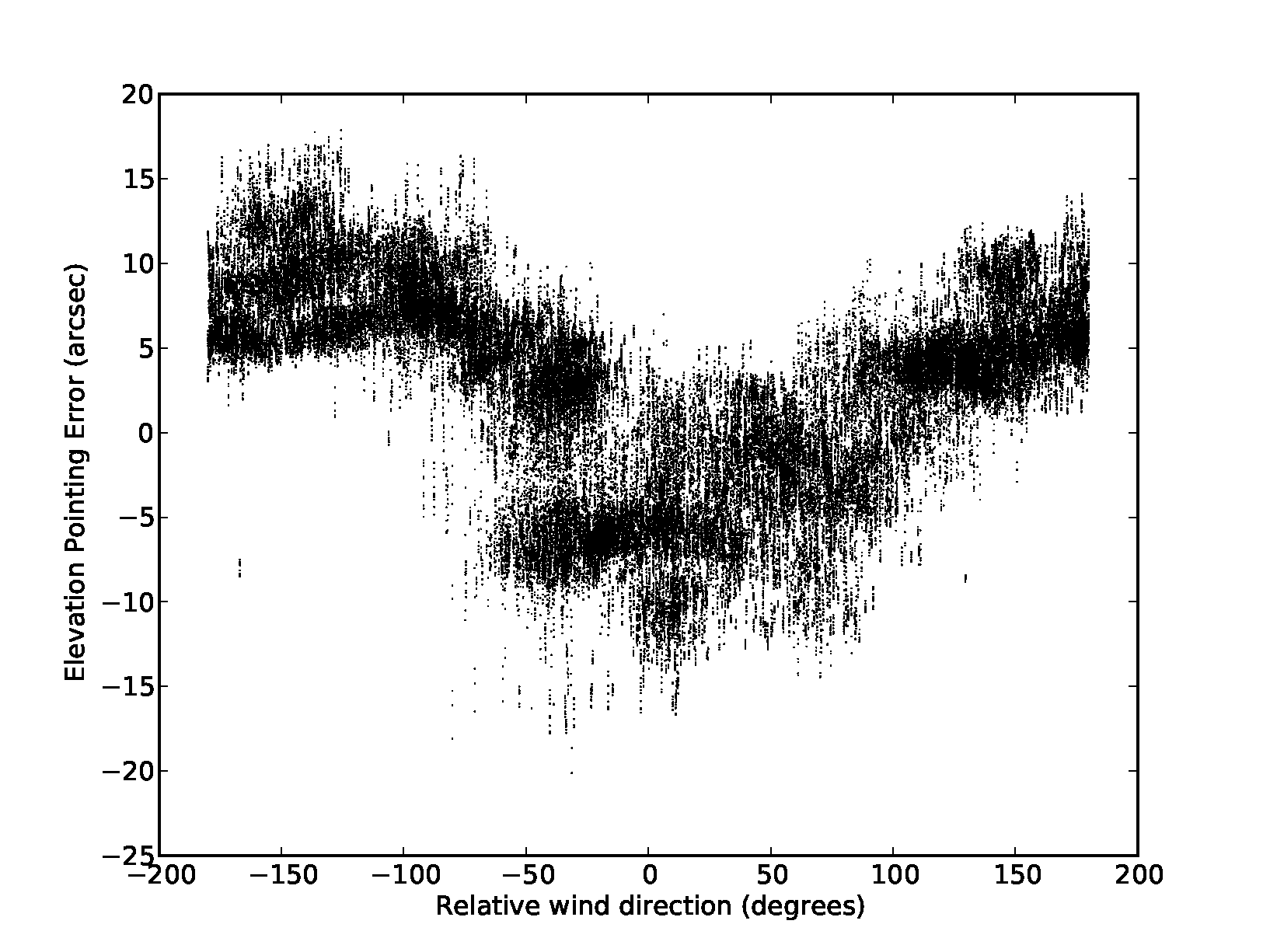}
\caption{Plots showing QD response to a change in wind direction
induced by rotating the telescope $360\arcdeg$ clockwise and
counterclockwise on 30 March 2010.  The average wind speed during each
rotation is 6-7 m~s$^{-1}$.  The wind direction in the above plots represents
measured wind direction minus telescope azimuth. In spite of substantial
fluctuations in wind direction during the hours of these rotations,
the pattern holds true.}
\label{fig:windvsdir}
\end{figure}

Various attempts to compare wind speed to QD-measured feedarm deflection have not shown an obvious correlation, although RMS QD pointing errors vary roughly with $v_\mathrm{wind}^2$.  Instead, we have discovered that the direction of the wind is actually the most relevant quantity.  On a windy day, we locked the telescope elevation axis at the survival elevation of 60 degrees, which is the position where the dish is closest to horizontal, and rotated the telescope $360\arcdeg$ about the azimuth axis in each direction.  The QD-inferred pointing error versus the wind direction relative to the telescope pointing direction (accounting for both wind shifts and the ongoing rotation of the telescope) is shown in Figure \ref{fig:windvsdir}.  The physical mechanism for this response is that the wind is blowing the feedarm forward along the wind vector, which results in a mostly sinusoidal signature in both elevation and cross-elevation pointing error, with a phase difference of $90\arcdeg$.  The feedarm response is twice as large in the cross-elevation direction as it is in the elevation direction.  We suspect that this effect is due to reduced structural stiffness in cross elevation rather a change in drag coefficient since the drag coefficient of a lattice structure such as the feedarm only varies by about 10\% with changes in angle of attack \citep[p. 440]{simiu1996}. The telescope dish can also provide a shielding effect when the wind is blowing from the front or back of the dish, and this effect has been observed in the aforementioned serendipitous dataset.  However, since the telescope was locked at survival elevation during the rotation experiment, any shielding effect should be minimized. This result provides important constraints on the common use of the offset pointing technique (in which the local pointing corrections are periodically measured by pointing on a quasar near on the sky to the science target \citep{condon2001}).  Clearly, under windy conditions, it is more important to choose a pointing calibrator nearby in azimuth rather than elevation to minimize this source of error due to the change in the projected angle to the wind force vector.  From these data, the pointing change incurred by a worst case $180\arcdeg$  slew can be as large as $50''$ in 6~m~s$^{-1}$ of wind.  

The cross-elevation direction component of the QD calibration was determined astronomically, so it could represent pointing errors from sources other than just the feedarm, such as coupled motion of the main dish.  In order to discriminate between the motion of the feedarm and the motion of the telescope as a whole, we used three three-axis accelerometers on the telescope: one on the ceiling of the Gregorian receiver cabin near the illuminator (accelerometer 1), one at the edge of the dish nearest the feedarm (accelerometer 4), and one immediately adjacent to the QD detector
(accelerometer 5).  Figure \ref{fig:gbtlayout} shows these locations.  The accelerometer data were processed to remove the signal of gravitational  deflection in a manner similar to that used for removing the signal from the QD data since they are all located on the tipping structure.  We then compared the magnitude of the accelerations at these three accelerometers under windy conditions and under deliberate feedarm excitation.  The result was the same in both cases: motion at accelerometer 1 was much greater than motion at accelerometer 4, which was greater than motion at accelerometer 5.  In other words, the telescope pointing errors are almost entirely due to the motion of the feedarm alone (represented by accelerometer 1), rather than the coupled motion of the dish (represented by accelerometers 4 and 5).

\section{Astronomical use with MUSTANG}

MUSTANG (MUltiplexed Squid Transition-edge sensor Array at Ninety
Gigahertz) is an imaging bolometer array used on the GBT at
3.3~mm/90~GHz \citep{mustang2008}.  It is currently the highest
frequency instrument in use on the GBT.  As such, it is the instrument
that would benefit the most from pointing improvements.  MUSTANG's
primary scan pattern is not a simple stare at a source, but rather a
daisy petal, billiard ball, or other moving scan.  As a result, the
data reduction pipeline for MUSTANG must use the timestream of the
position of the telescope as a critical input.  Therefore, there is
little need to correct for static or dynamic pointing errors in real
time.  So long as the errors are recorded and smaller than the field
of view of the array ($30''$), they can be corrected during the
data reduction stage.

\begin{figure}[H]
\centering
\resizebox{120mm}{!}{\plotone{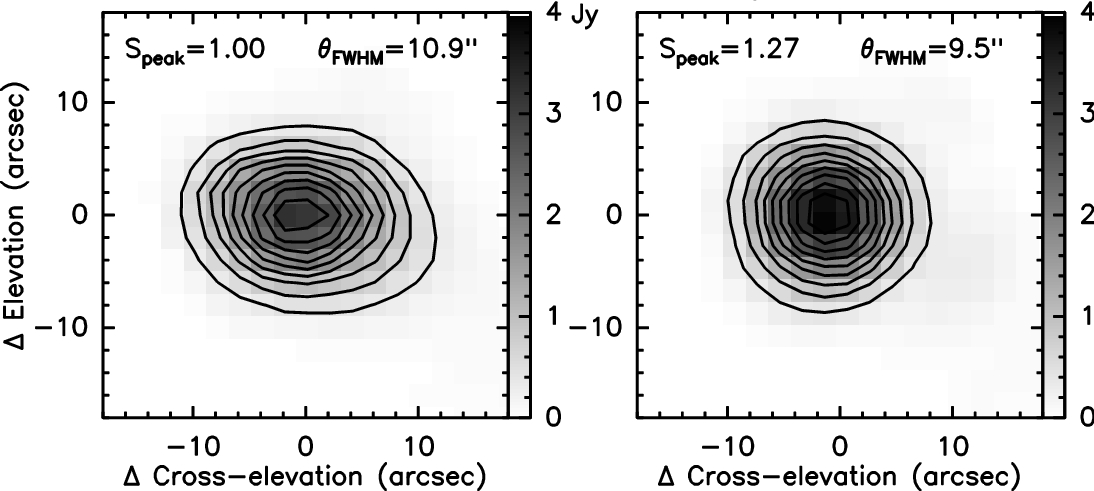}}
\caption{MUSTANG images of a bright quasar obtained on a windy ($\approx$ 6
  m~s$^{-1}$) night combining several scans with a total integration time of
  about 20 minutes from March 4, 2010.  The image on the left has had no QD correction
  applied.  The image on the right has had a QD correction applied
  using one median value to remove QD drift for all of the scans.  For
  reference, the typical size of a MUSTANG beam under ideal conditions is
  $9.0''$.}
\label{fig:globalmed}
\end{figure}

Such corrections have already been implemented into the MUSTANG data
reduction software written in IDL.  By default, the MUSTANG pipeline
takes QD data recorded during each scan, subtracts a median from the
QD data from that scan in order to remove drift in the QD, then
removes the QD-inferred pointing errors from the data stream.  More
advanced users can choose to compute a median that is derived and used
over several scans.  As a result, the influence of the wind can be
nearly completely removed.  Figure \ref{fig:globalmed} shows the
improvements the QD can make over the course of 20 minutes integration
time on a windy night.  The primary effect of wind is to smear out the
source in the cross-elevation direction.  When used properly, the QD
can remove a majority of the detrimental effects of observing under
windy conditions.

A precise figure is difficult to calculate since this experiment was done on a bright quasar on a windy night. Comparable data from a calm observing run is not available for direct comparison due to quasar variability, telescope efficiency variability, and other factors.  We assume that point source sensitivity goes as $\frac{P}{W^2}$, where $P$ is the peak of the Gaussian point-spread function (PSF) and $W$ is the width of the Gaussian PSF (i.e. proportional to flux density on the sky, brightness per unit area).  We do not know the true peak of the source, however.  We calculated an ideal peak based on the peak value required to make a Gaussian with the same total integrated flux of the windy Gaussian, but with an ideal FHWM, which is more predictable.  The range given here reflects different assumptions in the calm PSF, with 9 arcseconds being the absolute best and 9.3 arcseconds representing a more typical PSF observed under calm conditions. In wind speeds of 5m/s, we go from an uncorrected point source sensitivity of 47$-$53\% to a corrected sensitivity of 78$-$88\% compared to our assumed benign-weather sensitivity, for a recovery of 58$-$74\% of the sensitivity lost to wind.  Several scientific papers have been published using MUSTANG data processed with the QD correction as the authors have found that it improves the consistency of the results \citep{cotton2009,dicker2009,mason2010,korngut2010,shirley2010}.

\section{Other Astronomical Uses}

\begin{figure}[H]
\centering
\plottwo{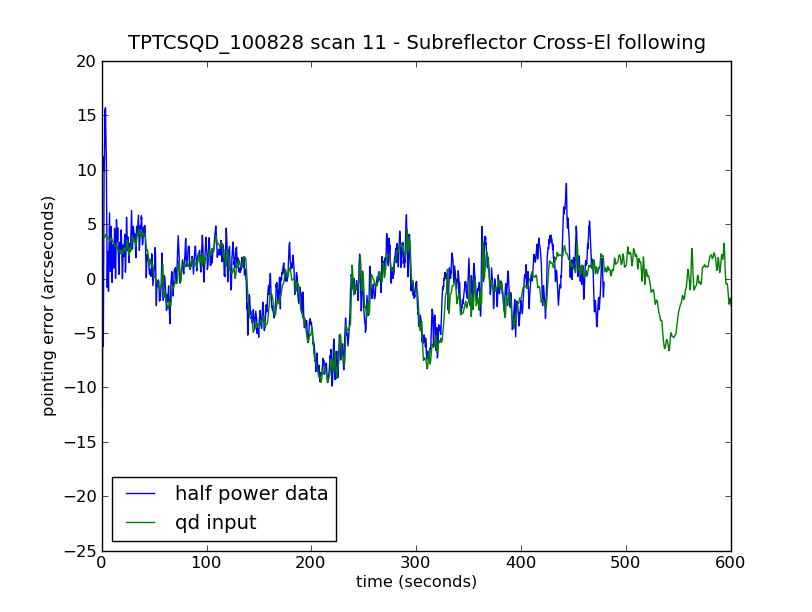}{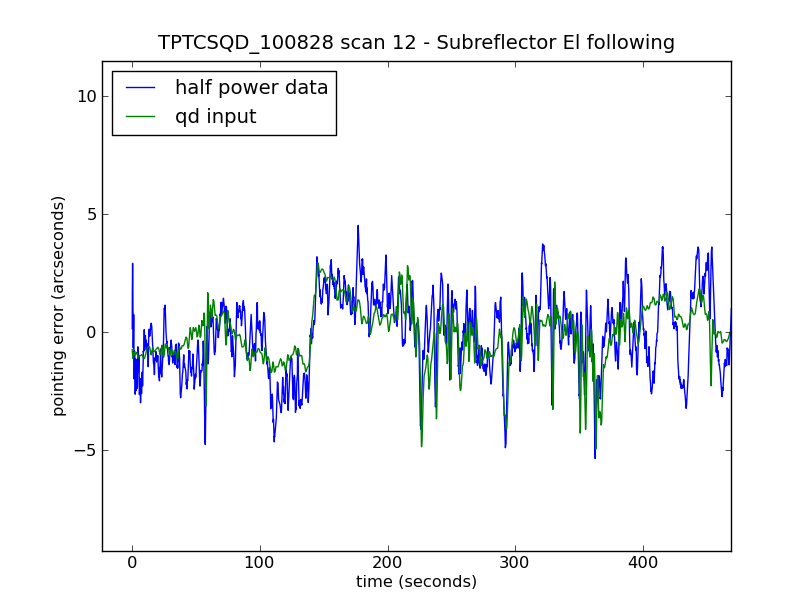}
\caption{An experiment demonstrating the potential usefulness of the
  QD to correct pointing in real-time.  The QD signal from a windy
  dataset is used as an input to the subreflector on a calm day.  The
  resulting pointing of the telescope is then measured during a
  half-power track on a quasar. Left panel) a cross-elevation
  half-power track.  Right panel) an elevation half-power track.  Both
  axes of the subreflector are easily able to follow the real QD data
  as an input trajectory and produce the expected pointing error.}
\label{fig:qdsubref}
\end{figure}

The QD can benefit observers using GBT instruments other than MUSTANG as well. One future project is to incorporate a flag in the observing data whenever the beam has moved significantly off-source according to the QD. Users can then either delete these time periods or attempt to correct the measured signal during these time periods.  Either method will improve the accuracy and consistency of the measurement even though the loss in signal-to-noise ratio cannot be recovered.  However, the ideal method of QD application would be to somehow apply corrections in real time while observing, to prevent a loss of sensitivity in the first place.  The subreflector servo system on the GBT provides this capability.  Recently, we performed an experiment of taking real QD data recorded under windy conditions and feeding it into the subreflector servo during a half-power track in calm conditions to demonstrate that the servo can follow input from the QD.  The resulting data shown in Figure \ref{fig:qdsubref} indicate that the telescope pointing was perturbed exactly as expected.  Thus, it is clear that the subreflector is capable of following realistic data from the QD. Since the bandwidth of the QD correction is only a few Hz, there is sufficient time (i.e. many milliseconds) to perform the small amount of signal processing and update accordingly the subreflector trajectory.  The only remaining task is to connect the real-time data path in the GBT M\&C system, which we expect to attempt in the near future.  After a suitable testing period, real-time corrections with the subreflector will made available as an option for high-frequency observers, thus eliminating the need for any corrections in the post processing if enabled.

This work was supported in part by the NRAO traineeship program.  We
are grateful to the NRAO staff in Green Bank (past and present) for
their many contributions to this research program.  The National Radio
Astronomy Observatory is a facility of the National Science Foundation
operated under cooperative agreement by Associated Universities, Inc.  We would also like to thank the anonymous referee for several useful suggestions for improving this paper.

\bibliographystyle{astron}
\bibliography{ms}

\end{document}